# Nonlinear down-conversion in a single quantum dot

B. Jonas[1,2,3], D. Heinze[1,2], E. Schöll[1,2,3], P. Kallert[1,2,3], T. Langer[1,2,3], S. Krehs[1,2,3], A. Widhalm[1,2,3], K. D. Jöns[1,2,3], D. Reuter[1,2,3], S. Schumacher[1,2,4,*], and A. Zrenner[1,2,3,†]

[1] Paderborn University, Physics Department, Warburger Straße 100, 33098 Paderborn, Germany
[2] Paderborn University, Center for Optoelectronics and Photonics Paderborn (CeOPP), Warburger Straße 100, 33098 Paderborn, Germany
[3] Paderborn University, Institute for Photonic Quantum Systems, Warburger Straße 100, 33098 Paderborn, Germany
[4] Wyant College of Optical Sciences, University of Arizona, Tucson, Arizona 85721, USA

*stefan.schumacher@upb.de
†artur.zrenner@upb.de

*Photonic quantum technologies[1], with applications in quantum communication, sensing as well as quantum simulation and computing, are on the verge of becoming commercially available. One crucial building block are tailored nanoscale integratable quantum light sources, matching the specific needs of use-cases. Several different approaches to realize solid-state quantum emitters[2] with high performance[3] have been pursued. However, the properties of the emitted single photons are always defined by the individual quantum light source and despite numerous quantum emitter tuning techniques[4–7], scalability is still a major challenge. Here we show an emitter-independent method to tailor and control the properties of the single photon emission. We demonstrate a laser-controlled down-conversion process from an excited state of a quantum three-level system[8]. Starting from a biexciton state, a tunable control laser field defines a virtual state in a stimulated process. From there, spontaneous emission to the ground state leads to optically controlled single photon emission. Based on this concept, we demonstrate energy tuning of the single photon emission with a control laser field. The nature of the involved quantum states furthermore provides a unique basis for the future control of polarization and bandwidth, as predicted by theory[9,10]. Our demonstration marks an important step towards tailored single photon emission from a photonic quantum system based on quantum optical principles.*

Today's quantum dot (QD) research and applications employ resonant excitation schemes, which give access to functionalities known from two-level systems. This is the basis for coherent state control and manipulation of excitonic states, which are important for photonic quantum technologies. Rabi rotations are the standard technique for the excitation of single photon emitters[11–13]. This approach has proven to be the best scheme to realize deterministic sources of indistinguishable photons[14], which are required for the exploitation of single photon nonlinearities, based on the Hong-Ou-Mandel effect (HOM) in photonic quantum computing and simulation[15,16]. For the generation of polarization entangled photon pairs, this scheme was extended to the resonant two-photon excitation (TPE) of the biexciton state $|B\rangle$. For vanishing fine-structure splitting of the exciton state $|X\rangle$, the sequential decay via two equivalent paths leads to polarization entanglement of the two emitted photons[17]. In the schemes described so far the spectral and polarization properties of the emitted photons are defined by properties of the quantum states, which are involved in the decay process. The quantum states can also be fine-tuned by applying an electric field[4], a magnetic field[5], or strain[6,18,19]. Also photonic methods have been employed in the past to control the emission of single quantum systems. A prominent example is the Λ-system, where Raman-like transitions via a virtual state lead to single photon emission that can be controlled by the spectral properties of the driving laser field[20]. In a different approach, a cavity has been used to steer the biexciton decay to a degenerate, spontaneous two-photon decay path[21].

In the present contribution we introduce and demonstrate a new concept, which enables the control of single photon emission from an individual semiconductor quantum dot by a nonlinear down-conversion process. Starting from a biexciton state $|B\rangle$, a tunable control-laser field defines a virtual state for a stimulated down-conversion (SDC). Related concepts, which are reminiscent to stimulated Raman scattering[22], have been studied already in atomic systems[23–25].

The SDC process in a single QD is schematically outlined in Fig. 1. In an initial step, a resonant TPE of the biexciton state $|B\rangle$ with energy $E_B$ is performed. By driving the system with a tunable, off-resonant control-laser field, a virtual state for the SDC process is defined. From there, spontaneous emission to the ground state $|G\rangle$ leads to optically controlled single photon emission. The SDC-emission energy $E_{SDC}$ is therefore correlated with the control-laser energy $E_C$ and the simple energy-conservation $E_B=E_C+E_{SDC}$ holds.

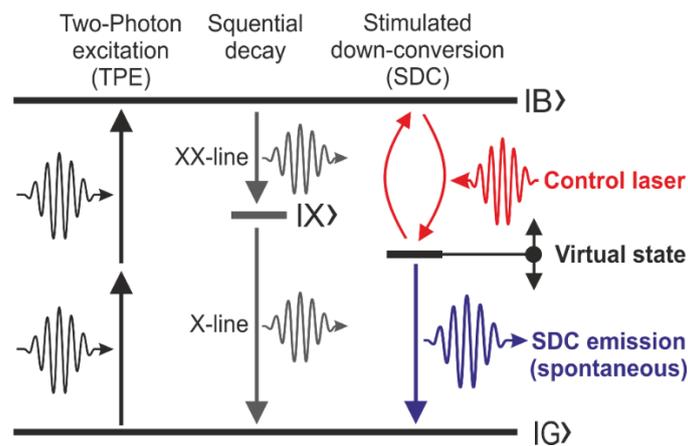

**Fig. 1| Excitation and down-conversion scheme.** Two-photon excitation from the ground state $|G\rangle$ to the biexciton state $|B\rangle$ and stimulated down-conversion (SDC) to a virtual state by a control laser field. The energy of the spontaneous SDC emission is defined by the energies of the state $|B\rangle$ and the control-laser energy and detuned from the cascaded decay via the exciton state $|X\rangle$.

The initial biexciton state is composed of two electrons and two holes, each with antiparallel spins, in the s-shell of the QD. Therefore, the initial state $|B\rangle$ and the final state $|G\rangle$ are both spin-zero states. Consequently, the spins of the involved photons (control and SDC), must add up to zero. Hence, the energy and polarization of the single photon emission can be controlled by the energy and polarization of the classical control-laser field.

This optical control of tailored single photon emission has been intensively studied in theory, with a special focus on timing and selection rules[9], spectral properties of the resulting emission[10], and methodological parallels to the description of Λ-systems[8].

In our experimental realization we use (In,Ga)As QDs grown by molecular beam epitaxy (MBE), which are embedded in a GaAs p-i-n photodiode (see Fig. 2c) to allow for Stark tuning of the transition energies. The QDs are placed in the center of the 320 nm thick intrinsic layer. The areal density of the embedded (In,Ga)As-QDs is in the range of 1 QD/µm², low enough to allow for single QD spectroscopy. The buried n-GaAs region has a thickness of 110 nm and a Silicon doping concentration of $2\times10^{18}$ cm$^{-3}$. The Carbon doped p-GaAs region consist of a 95 nm thick layer with a doping concentration of $4\times10^{18}$ cm$^{-3}$, followed by a 15 nm thick surface layer with a doping concentration of $2\times10^{19}$ cm$^{-3}$. From this material, large area quantum dot photodiodes (550 x 700 µm²) are processed. On top of those, super-hemispherical solid immersion lenses (diameter 500 µm) are placed to enhance the detection efficiency and light-matter interaction (see Fig. 2b).

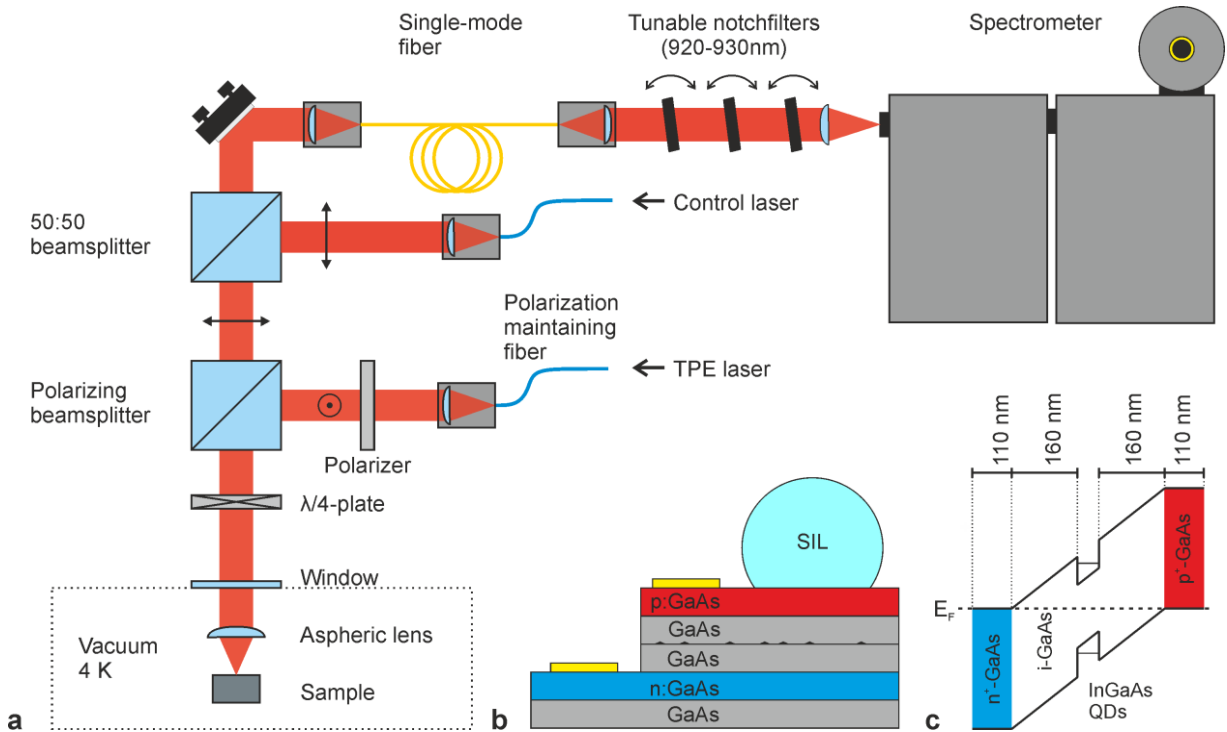

**Fig. 2 | Experimental setup and sample structure. a**, Optical setup for the down-conversion experiment (see also text). **b,** p-i-n QD mesa-diode with attached solid-immersion lens (SIL). **c,** Band diagram of the MBE-grown p-i-n structure with indicated thickness of the individual layers.

The sample is placed in a low temperature microscope setup, which is cooled down to T = 4.2 K (see Fig. 2a). An aspheric lens with NA = 0.68 is used to focus the laser light onto the specimen and to collect its emission in a confocal geometry. As excitation sources we use two independently tunable continuous wave (cw) lasers, both fiber-coupled to the head of the microscope: An external cavity diode laser for the TPE and a Ti:Sapphire-laser in cw-mode as control laser. The emission of both lasers is filtered with tunable band-pass filters with a bandwidth of 0.45 nm to suppress sideband emission (not shown in Fig. 2a). The QD emission is collected with a single-mode optical fiber and sent to a spectrometer. In order to detect the QD emission, the stray light resulting from the two lasers must be suppressed. For the TPE laser, this is realized by a combination of a polarization suppression scheme based on the setup published by Kuhlmann et al.[26] and a tunable notch-filter with a bandwidth of 0.7 nm in front of the spectrometer. Because of the selection rules, the control laser cannot be suppressed by cross-polarization. Therefore, we use spectral filtering with two additional notch-filters. This setup finally allows for s-polarized excitation by the TPE laser, a p-polarized control laser and detection of the p-polarized contribution of the QD emission.

The starting point is the initial state preparation for the optical down-conversion experiment illustrated in Fig. 1. Biexciton preparation can be achieved with a variety of excitation schemes[27–29]. In this work we use phonon-assisted two-photon excitation. This process has the advantage that it is very robust against detuning and changes of the laser power[30,31]. Applying a bias voltage ($V_B$) to the diode structure tunes the emission energies of the QD due to the Stark effect. Experimental data for a bias voltage controlled tuning range of about 2 meV is displayed in Fig 3. The TPE laser is tuned to 1341.17 meV and set to an excitation power of 12 mW to achieve saturated excitation conditions[4,24,25].

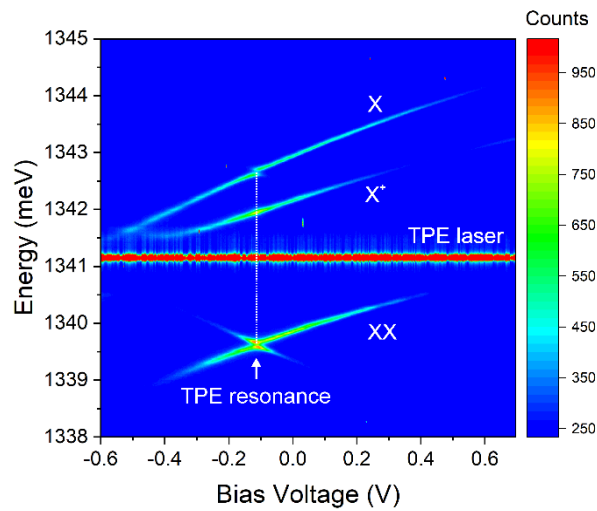

**Fig. 3 | Two-photon excitation of the biexciton.** Bias voltage dependent results for the phonon assisted two-photon excitation of the biexciton by a TPE laser without control laser. The sequential decay of the biexciton leads to the XX- and X-emission lines. The emission of the positive trion-line $X^+$ appears statistically in the biased diode structure. At the TPE-resonance, the avoided crossings of the XX- and X-lines are caused by the dressing of the $|B\rangle$- and $|G\rangle$-states ($T_{int}$=1 s).

Due to the polarization suppression scheme we only detect light with a polarization perpendicular to the TPE laser. The measurement shows the emission of the XX-/X-cascade and an emission of the positive trion ($X^+$), which statistically appears in the biased diode structure. At $V_B$ = -0.12 V the laser is in resonance with the direct two-photon transition and we achieve direct two-photon excitation without the need of phonons. As a result of the high excitation power, an avoided crossing in the X- and XX-line is observed, which is the result of the dressing of the $|B\rangle$- and $|G\rangle$-states[32–34].

The phonon assisted TPE provides the preparation of the initial state for our down-conversion experiment. For this we introduce a control laser, which is tuned close to the energy of either the XX- or X-line. Due to the selection rules for the down-conversion process, the emitted photons have the same linear polarization as the control laser. In our setup with polarization suppression for the TPE-laser, the polarization of the control laser has to be perpendicular to the TPE laser. Therefore, this laser can only be suppressed by spectral filtering, which we realize with two tunable notch-filters.

It is expected that resonance enhancement in the efficiency of the SDC can be observed when the laser induced virtual state is near the $|X\rangle$-state. Based on this consideration and the conservation of energy, there are two favorable scenarios for the demonstration of SDC. Here, the control laser is either tuned close to the X- or the XX-line, resulting in SDC emission in the vicinity of the XX- or X-line respectively (see Fig. 4). To ensure an unambiguous identification of the down-converted emission we make use of the Stark effect tuning characteristics of the exciton states in the QD. In Fig. 4 we schematically show the tuning characteristics of the single exciton state $|X\rangle$ and the biexciton state $|B\rangle$. The latter has a Stark shift roughly twice as large as compared to the shift of the $|X\rangle$ state. If we switch on a control laser field with a photon energy $E_C$, there are two equivalent options for the formation of a virtual state. In Fig. 4 a, we have illustrated a scenario, where a virtual state at $E=E_G+E_C$ forms the final state for a spontaneous SDC decay from the biexciton state $|B\rangle$. An alternative scenario is illustrated in Fig. 4 b, where a virtual state at $E=E_B-E_C$ forms the initial state for a spontaneous SDC decay to the ground state $|G\rangle$. For both scenarios the SDC decay with energy $E_{SDC}$ appears with a Stark shift, which corresponds to the shift of the biexciton state $|B\rangle$. The SDC-emission can therefore be identified by its unique Stark shift fingerprint, which deviates from the behavior of the allowed and competing X- and XX-lines. The characteristic Stark-shift of the SDC-emission is also a direct consequence of the expected conservation of energy: $E_C+E_{SDC}=E_B$. Following this strategy, we analyzed the down-conversion process in bias voltage dependent experiments.

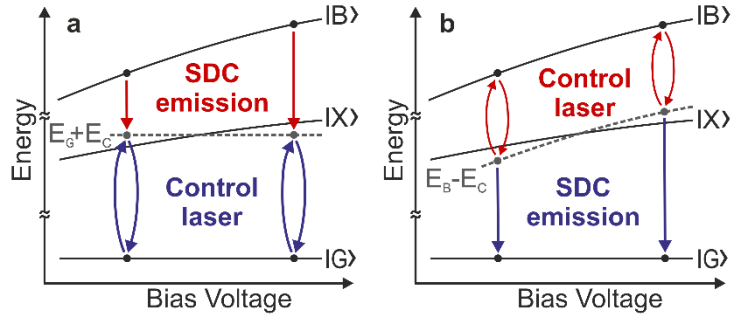

**Fig. 4| Stark shift behavior of the SDC-emission. a**, A control laser energy higher than the TPE laser energy results in SDC-emission from $|B\rangle$ to the virtual state at $E=E_G+E_C$. **b,** A control laser energy lower than the TPE laser energy results in SDC emission from the virtual state at $E=E_B-E_C$ to $|G\rangle$. In both cases the Stark-shift of the SDC emission is given by the Stark-shift of the $|B\rangle$ state.

In Fig. 5 we present our results, which conclusively demonstrate laser-controlled down-conversion in a single QD. For this we applied control laser energies above (Fig. 5 a) and below (Fig. 5 b) the TPE laser energy. For both cases we observe the predicted down-conversion process (labeled SDC), in each case on the complementary energy side and also with the expected magnitude of the Stark shift, given by the tuning behavior of the biexciton state $|B\rangle$ (sum of the slopes of the X- and XX-lines).

When the X-line is tuned into resonance with the control laser (see Fig. 5 a for a bias voltage of $V_B = 0.2$ V), the single exciton state $|X\rangle$ and ground state $|G\rangle$ are split (dressed) and form a Mollow-triplet[35] (see level scheme in Fig. 6 a). This interpretation is supported by the comparison with our theoretical results shown in Fig. 5 c. Details on the theoretical approach are given further below and in the appendix. In this case the SDC process takes place between the $|B\rangle$ and a dressed $|X\rangle$ state (split) and the XX-line shows an avoided crossing with the SDC-emission. The comparison of our experimental results with the simulation shown in Fig. 5 c, shows very convincing agreement and all features can be explained with our model.

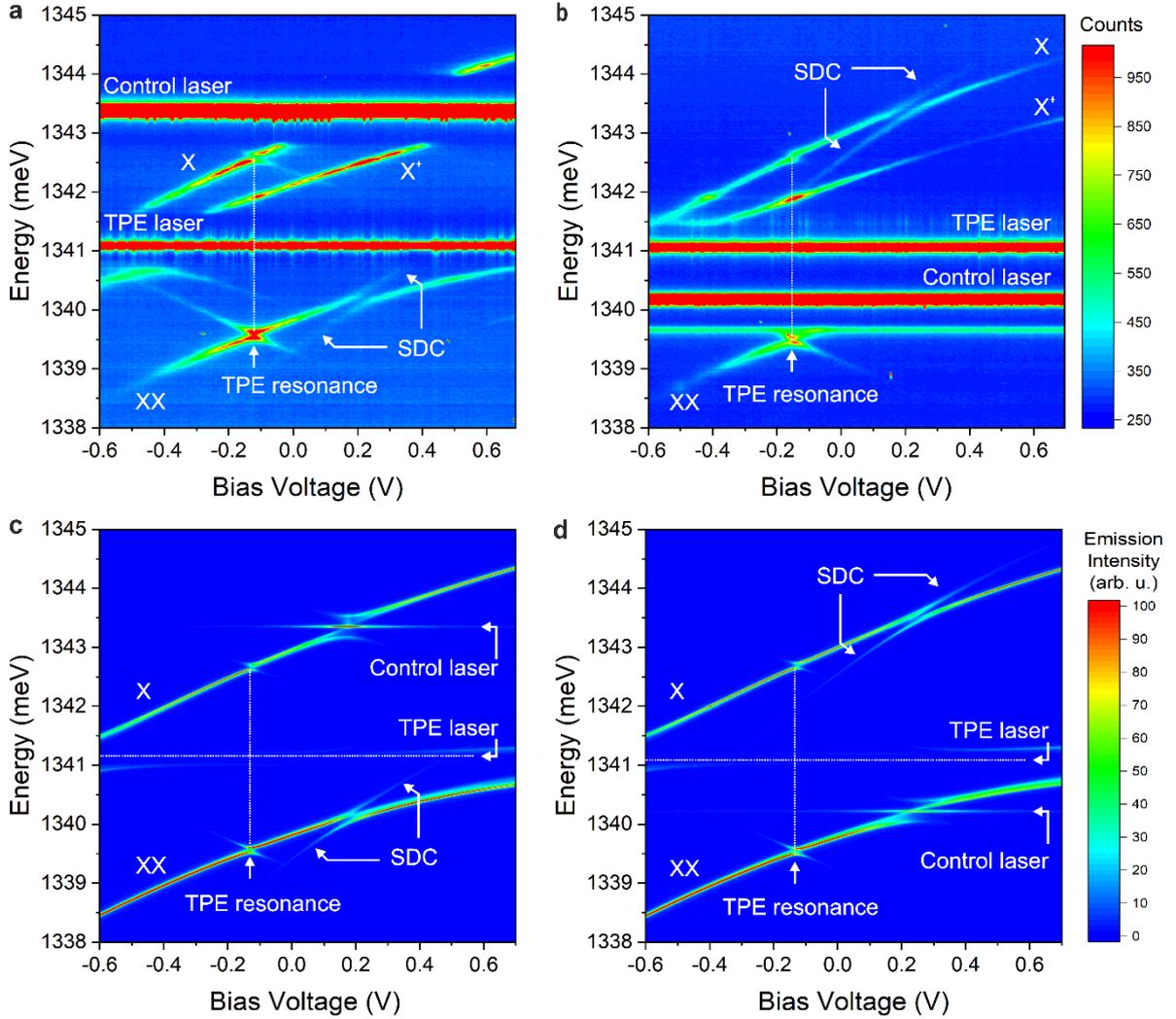

**Fig. 5 | Stimulated down-conversion (SDC) in a single quantum dot. a,** Experiment with blue-shifted control laser: The SDC emission appears on the low-energy side with the expected Stark shift and an avoided crossing with the XX-line *(P$_{TPE}$=10 mW, P$_{Control}$=1.5 mW, T$_{int}$=2 s)*. **b,** Experiment with red-shifted control laser: The SDC emission appears on the high-energy side with the expected Stark shift and an avoided crossing with the X-line. **c,** Theoretical results for case **a**, taking into account experimental resonance energies and Stark shifts. **d,** Theoretical results for case **b**.

Please note that notch filters are used in the experiment to suppress stray light from the TPE- and control laser. In a narrow range around their center energies (about ±0.4 meV) the QD emission is therefore blocked.

The theoretical results shown in Fig. 5 c and d are based on a density matrix description of the system including the relevant electronic configurations $|G\rangle$, $|X\rangle$, and $|B\rangle$ and classical coherent light fields. The von-Neumann equation of motion for the system density operator is solved in matrix representation including pure dephasing and radiative losses. To approximate the cw coherent excitation of the experiments, long Gaussian pulses are used. An additional incoherent source is included to enable biexciton excitation for the case when the two-photon resonance condition is not met by the laser source. Emission spectra are obtained from the time-integrated Fourier transform of the first order correlation function $g^{(1)}(t,\tau)$ as detailed in the appendix.

For a control laser energy below the TPE laser energy (Fig. 5 b), the SDC-emission and the avoided crossing is experimentally observed on the X-line at $V_B$ = 0.2 V. In analogy to the previous case, the Mollow-triplet appears now on the XX-line as observed in the theoretical results in Fig. 5 d. The corresponding level scheme is shown in Fig. 6 b.

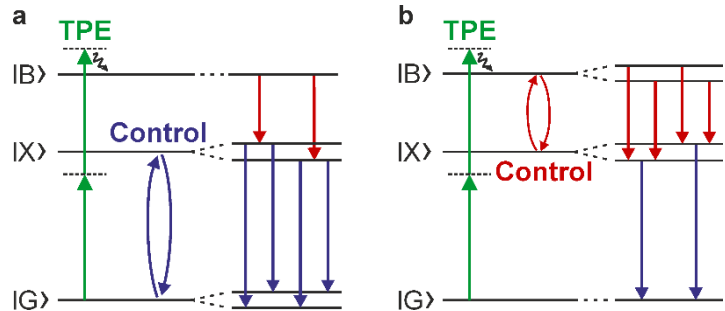

**Fig. 6 | Energy levels for resonant control laser. a,** The control laser drives the $|X\rangle$ - $|B\rangle$-transition, creating dressed states and forming a Mollow triplet. The $|B\rangle$ - $|X\rangle$-transition is split in two lines. **b,** The control laser drives the $|B\rangle$ - $|X\rangle$-transition, creating dressed states and a Mollow triplet. The $|X\rangle$-$|G\rangle$-transition is split in two lines.

Our experimental and theoretical results demonstrate in an unambiguous way stimulated down-conversion in a single QD. We conclude this from the characteristic Stark-effect shift of the SDC-emission, which is a consequence of the conservation of energy. Conceptually, this mechanism lays the foundations for a new generation of nonlinear optical devices on the single quantum level. The down-conversion process works best in close energetic proximity of the $|X\rangle$ state. With a tuning range up to 0.5 meV, the down-converted emission is observable also far from the avoided crossing with the $|X\rangle$ state. With an exceptionally good agreement between theory and experiment, the theory provides additional insights that complement the available experimental evidence. This is true in particular for the appearance of the Mollow-triplet, that is not observable in the experiment, as a result of the required stray-light suppression.

For future applications we expect that the efficiency and tuning range of the SDC-process can be vastly improved by the employment of cavities with high Purcell-factor. This will enhance the SDC-emission and suppress the competing emission cascade via the $|X\rangle$ state[9]. Theory further predicts polarization control of the single photon emission by adjusting the polarization of the control laser. Pulsed two-photon excitation allows for the deterministic preparation of the $|B\rangle$-state. Control over the timing and the spectral characteristics of the single photon emission can be gained by tailored control laser pulses. Overall, our results open entirely new pathways for the optical control of quantum emitters by nonlinear concepts.

## Acknowledgements
This work was supported by the Deutsche Forschungsgemeinschaft (DFG) through the transregional collaborative research center TRR142 (grant No. 231447078, project A03), by the Heisenberg program (grant No. 270619725), by the BMBF via the Q.Link.X Project No. 16KIS0863, and by the Paderborn Center for Parallel Computing, PC[2]. KJ further acknowledges funding from the European Union's Horizon 2020 research and innovation program under grant agreement No. 820423 (S2QUIP) and No. 899814 (Qurope).

# Appendix

To model the optical excitation of the quantum dot system we include four electronic configurations - the ground state G, two exciton finestructure states X$_H$ and X$_V$ as well as the biexciton B - with the Hamiltonian

$$H_0 = E_G|G\rangle\langle G| + E_H|X_H\rangle\langle X_H| + E_V|X_H\rangle\langle X_H| + E_B|B\rangle\langle B|$$

We adjust the voltage dependent (bi-)exciton energies to the experimental data shown in Fig. 3. Corresponding to the experimental situation, two lasers couple adjoining electronic states with the selection rules

$$H_{int} = \sum_{i=H,V} ((|G\rangle\langle X_i|\Omega_i^*(t) + |X_i\rangle\langle B|\Omega_i^*(t)) + h.c.)$$

with $\Omega_H$ and $\Omega_V$ being the amplitude of the TPE and control laser, respectively. Here, we assume long Gaussian pulses on the time scale of the losses present in the system to mimic the response of the cw - excited quantum dot as

$$\Omega_i(t) = \frac{\Omega_i^0}{\hbar} exp(-\frac{(t-t_0^i)^2}{2\Sigma^2} + i\omega_i t)$$

with $\Omega_H^0 = 288.7\ \mu eV$, $\Omega_V^0 = 103.1\ \mu eV$, $t_0^H = 200\ ps$, $t_0^V = 300\ ps$, and $\Sigma = 100\ ps$. The pulse frequencies $\omega_i$ are chosen according to the experimental data in Fig. 5. For comparison with the measured spectra, the theoretical emission intensity[36] is calculated with respect to the electronic quantum dot operator $\sigma_V = |G\rangle\langle X_V| + |X_V\rangle\langle B|$ in the same linear mode as the control pulse[37] as

$$S_V(\omega) = \Re \int_0^{t_0^V} dt \int_0^{t_0^V-t} d\tau \langle \sigma_V^\dagger(t)\sigma_V(t+\tau)\rangle e^{i\omega\tau} e^{-\frac{\Gamma}{2}(t_0^V-t)} e^{-\frac{\Gamma}{2}(t_0^V-t-\tau)}$$

with $\hbar\Gamma = 4\ \mu eV$. The integrated emission intensity is limited to the time window with upper limit $t_0^V$ to avoid contributions from the spontaneous emission not driven by the pulses. We evaluate the two-time expectation values using the quantum regression theorem[38] and taking the trace with the density matrix of the quantum dot system $\rho_s$ which obeys[9,37]

$$\frac{\partial \rho_s}{\partial t} = \frac{1}{i\hbar}[H_0 + H_{int}, \rho_s] - \frac{\gamma_{pure}}{2} \sum_{\chi,\chi';\chi\neq\chi'} |\chi\rangle\langle\chi|\rho_s|\chi'\rangle\langle\chi'|$$
$$+ \gamma_{rad} \sum_{i=X_H,X_V} (\mathcal{L}_{|G\rangle\langle i|} + \mathcal{L}_{|i\rangle\langle B|})(\rho_s) + P_{incoh.}\mathcal{L}_{|B\rangle\langle G|}(\rho_s)$$

where $\chi,\chi' \in \{G, X_H, X_V, B\}$ and $\mathcal{L}_\sigma(\rho_s) = (2\sigma\rho_s\sigma^\dagger - \sigma^\dagger\sigma\rho_s - \rho_s\sigma^\dagger\sigma)$. We choose the pure dephasing rate $\hbar\gamma_{pure} = 4\ \mu eV$, a radiative decay $\hbar\gamma_{rad} = 4\ \mu eV$, and an incoherent pump $\hbar P_{incoh.} = 4\ \mu eV$. In addition to the light pulse, the biexciton to ground state transition is pumped incoherently by the last term to account for biexciton generation even if the driving lasers are off-resonant to a (cascaded) two-photon excitation scheme. For efficient evaluation, these calculations are performed in the interaction picture such that the relatively large carrier frequencies in $H_0$ do not have to be treated explicitly numerically.